\documentclass{aastex63}
\begin{document}

\title{How do Spitzer IRAC Fluxes Compare to HST CALSPEC?}

\author[0000-0001-9806-0551]{Ralph C. Bohlin}
\affiliation{Space Telescope Science Institute, 3700 San Martin Drive,
Baltimore,  MD 21218, USA; bohlin@stsci.edu}

\author[0000-0002-2413-5976]{Jessica E. Krick}
\affiliation{IPAC, MC 330-6, Caltech, 1200 E. California Blvd., Pasadena, CA
91125, USA; jkrick@caltech.edu}

\author[0000-0001-5340-6774]{Karl D. Gordon}
\affiliation{Space Telescope Science Institute, 3700 San Martin Drive,
Baltimore,  MD 21218, USA; kgordon@stsci.edu}

\author[0000-0001-8816-236X]{Ivan Hubeny}
\affiliation{The University of Arizona, Steward Observatory, 933 North Cherry
Avenue, Tucson, AZ 85719, USA; ihubeny.astr@gmail.com }

\begin{abstract} 

An accurate tabulation of stellar brightness in physical units is essential for 
a multitude of scientific endeavors. The HST/CALSPEC database of flux standards
contains many stars with spectral coverage in the 0.115--1 \micron\ range with
some extensions to longer wavelengths of 1.7 or 2.5 \micron. Modeled flux
distributions to 32 \micron\ for calibration of JWST complement the shorter
wavelength HST measurements. Understanding the differences between IRAC
observations and CALSPEC models is important for science that uses IR fluxes
from multiple  instruments, including JWST. The absolute flux of Spitzer
IRAC photometry at 3.6--8 \micron\ agrees with CALSPEC synthetic photometry to
1\% for the three prime HST standards G191B2B, GD153, and GD71. For a set of
17--22 A-star standards, the average IRAC difference rises from agreement at 3.6
\micron\ to 3.4 $\pm$0.1\% brighter than CALSPEC at 8 \micron. For a smaller set
of G-stars, the average of the IRAC photometry falls below CALSPEC by as much as
3.7 $\pm$0.3\% for IRAC1, while one G-star, P330E, is consistent with the A-star
ensemble of IRAC/CALSPEC ratios. \end{abstract}


\section{Introduction}			

Precise absolute flux calibration of astronomical spectra is crucial for
understanding the nature of cosmic sources. For example, the most precise flux
determinations are essential for the interpretion of the supernova data that
measure the accelerating expansion rate of the universe
\citep{scolnic14,stubbs15} and for understanding the nature of exoplanet host
stars \citep{tayar2020}. White dwarf (WD) model atmosphere calculations for the
three
CALSPEC\footnote{http://www.stsci.edu/hst/instrumentation/reference-data-for-calibration-and-tools/astronomical-catalogs/calspec}
primary WD standards, G191B2B,  GD153, and GD71, define the shape of the
instrumental flux calibrations and of the spectral energy distributions (SEDs),
i.e. flux (or flux density) as a function of wavelength from 0.115 to 2.5
\micron\ for the instruments on the Hubble Space Telescope (HST) and for the
CALSPEC flux scale \citep{bohlinetal14, bohlin2020}, while the absolute flux
levels depend on SI-traceable measurements of Sirius and Vega \citep{bohlin14,
bohlin2020}. CALSPEC is the database of primary and secondary 
spectrophotometric standard stars used for calibration of HST, James Webb Space
Telescope (JWST), Gaia, and other ground- and space-based instrumentation.

Comparisons between CALSPEC and Spitzer Space Telescope Infrared Array Camera
(IRAC) \citep{fazio2004} photometric fluxes at 3.6 (IRAC1), 4.5 (IRAC2), 5.8
(IRAC3), and 8.0 (IRAC4) \micron\ \citep{krick2021} measure the relative
precision among different spectral categories and provide additional constraints
on the CALSPEC infrared (IR) spectral energy distributions (SEDs), which are
defined by model atmosphere fluxes fitted to HST observations at shorter
wavelengths. These CALSPEC SEDs are the main source of standard stars that are
required for the flux calibration of the JWST \citep{gordon2022}. Currently,
there are no IR constraints on the CALSPEC models beyond 2.5 \micron\ when
NICMOS spectra exist, beyond 1.7 \micron\ for WFC3 IR grism spectra, or beyond 1
\micron\ whenever only STIS SEDs determine the long wavelength limit of the
observational data. Table~\ref{table:teff} defines the models in terms of the
$\chi^2$ revised fits for the O, B, A, and G stars over four parameters: Teff,
log g, log Z, and E(B-V), as described by \citet{bohlin2017}. The models for the
three primary WDs remain the same as in \citet{bohlin2020}. For three stars 16
Cyg B, 18 Sco, and HD159222, \citet{kovtyukh2003} confirms the
Table~\ref{table:teff} Teff effective temperatures to a worst case of 69~K for
18 Sco by an absorption line analysis.

The purpose of this paper is to compare the CALSPEC SEDs to the expanded and
reanalyzed IRAC photometry of \citet{krick2021}. Our results use the final
Spitzer/IRAC absolute flux calibration of \citet{carey2012}. The Spitzer
spacecraft started nominal operations on 2003 December 1 and continued until the
cryogen was exhausted on 2009 May 15. The  two shorter wavelength IRAC channels
at 3.6 and 4.5 \micron\ resumed operations for the warm mission starting on 2009
July 27 and continuing until 2020 January. No data are included in our analyses
from the 2009.37--2009.72 (2009 May 15--2009 Sep 19) warmup period or before
2003.90.

Section 2 presents our methodology and Section 3 details the data
analysis, including refined IRAC photometry corrections. Section 4 presents our
new NLTE model grid that replaces the BOSZ LTE grid \citep{bohlin2017}
for the extrapolation of the CALSPEC O and B star SEDs to 32 \micron. Section 5
discusses the rejected stars and variability, while Section 6 summarizes the
CALSPEC/IRAC comparison with our expanded  sample size of IRAC of photometry and
critiques the models that will define the JWST flux calibration.

\begin{deluxetable}{lccccc}		
\tabletypesize{\scriptsize}
\tablewidth{0pt}
\tablecolumns{6}
\tablecaption{\label{table:teff} Parameters of the Model Fits}
\tablehead{
\colhead{Star} &\colhead{$T_\mathrm{eff}$} &\colhead{$\log g$}
&\colhead{$[M/H]$} &\colhead{E(B-V)} & $\chi^2$}
\startdata
  & OB-stars &  &  &  & \\
      10LAC &34130 & 4.30 &-0.22 & 0.085 & 3.981   \\
     LAMLEP &29270 & 4.35 &-0.26 & 0.016 & 2.881   \\
      MUCOL &33390 & 4.40 &-0.23 & 0.015 & 3.053   \\
     ETAUMA &17500 & 4.50 &-0.12 & 0.000 & 3.245   \\
  & A-stars &  &  &  & \\     
     109VIR & 9760 & 3.55 &-0.07 & 0.022 & 1.893   \\
    1732526\tablenotemark{a} & 8660 & 4.10 &-0.42 & 0.037 & 2.917   \\
    1743045\tablenotemark{a} & 7330 & 3.45 &-0.43 & 0.004 & 1.110   \\
    1757132\tablenotemark{a} & 7400 & 3.45 & 0.00 & 0.001 & 0.850   \\
    1802271\tablenotemark{a} & 9060 & 4.05 &-0.49 & 0.017 & 0.843   \\
    1805292\tablenotemark{a} & 8580 & 4.00 &-0.11 & 0.033 & 0.730   \\ 
    1808347\tablenotemark{a} & 7850 & 3.75 &-0.90 & 0.017 & 2.835   \\ 
    1812095\tablenotemark{a} & 7800 & 3.60 & 0.12 & 0.003 & 0.459   \\ 
  BD60D1753\tablenotemark{a} & 9420 & 3.80 &-0.02 & 0.013 & 0.600   \\ 
     DELUMI & 9160 & 3.65 &-0.23 & 0.005 & 1.025   \\ 
    ETA1DOR &10160 & 4.00 &-0.54 & 0.000 & 0.623   \\ 
   HD101452 & 7400 & 3.80 & 0.16 & 0.021 & 0.701   \\ 
   HD116405 &10820 & 4.00 &-0.33 & 0.000 & 0.620   \\ 
   HD128998 & 9560 & 3.65 &-0.49 & 0.000 & 0.992   \\
    HD14943 & 7900 & 3.85 & 0.03 & 0.003 & 0.635   \\ 
   HD158485 & 8630 & 4.20 &-0.33 & 0.047 & 1.523   \\ 
   HD163466 & 7890 & 3.65 &-0.34 & 0.020 & 1.943   \\ 
   HD165459\tablenotemark{a} & 8580 & 4.20 & 0.05 & 0.022 & 0.555   \\ 
   HD180609 & 8550 & 3.95 &-0.46 & 0.034 & 0.604   \\ 
     HD2811 & 7960 & 3.55 &-0.36 & 0.021 & 2.424   \\ 
    HD37725\tablenotemark{a} & 8410 & 4.30 &-0.10 & 0.046 & 1.440   \\ 
    HD55677 & 8880 & 3.80 &-0.90 & 0.048 & 1.797   \\
  & G-star solar analogs &  &  &  & \\
     16CYGB & 5710 & 3.70 & 0.05 & 0.002 & 0.117   \\
      18SCO & 5730 & 3.35 &-0.12 & 0.000 & 0.272   \\ 
   HD106252 & 5850 & 4.05 &-0.08 & 0.001 & 0.145   \\ 
   HD159222 & 5820 & 3.75 & 0.13 & 0.000 & 0.152   \\ 
   HD205905 & 5860 & 3.85 & 0.06 & 0.001 & 0.170   \\ 
    HD37962 & 5700 & 3.75 &-0.22 & 0.000 & 0.230   \\ 
    HD38949 & 6000 & 4.35 &-0.11 & 0.000 & 0.083   \\ 
      P330E\tablenotemark{a} & 5830 & 4.90 &-0.21 & 0.028 & 0.360   \\ 
\enddata
\tablenotetext{a}{WFC3 and/or NICMOS grism data extend the observed SED longward
of 1~\micron.}
\tablecomments{Results from fitting model atmospheres to the observed stellar
SEDs using $\chi^2$ fitting with our new \textsc{tlusty} NLTE grid for the 
OB-stars and the BOSZ LTE grid \citep{bohlin2017} for A and G-stars. The
parameters of the fit for each star are the effective temperature
$T_\mathrm{eff}$, the surface gravity $\log g$, the metallicity [M/H]. the
interstellar reddening E(B-V), and the reduced chi-square quality of the fit
$\chi^2$.}
\end{deluxetable}

\section{Methodology}			

To compare the broadband Spitzer IRAC mean flux values $\langle F_{\nu}\rangle$
in mJy with the CALSPEC mean flux  $F_{\lambda}$ in
$erg~cm^{-2}~s^{-1}~\AA^{-1}$, the proper conversions must be applied.
The mean flux over the filter bandpass \citep{bohlinetal14} is
\begin{equation}\langle F_{\lambda}\rangle={\int F_\lambda~\lambda~R~d\lambda
\over \int \lambda~R~d\lambda}~,\label{eql}\end{equation} where $R$ is the
system fractional throughput as a function of wavelength from \citet{krick2021}.
The relation between the mean flux in frequency and wavelength units is 
\begin{equation}\langle F_{\lambda}\rangle~\lambda_p=\langle
F_{\nu}\rangle~\nu_p~,\label{eq2}\end{equation} where $\lambda_p$ and $\nu_p$
are the  pivot-wavelength and pivot-frequency. 
\begin{equation}\lambda_p=\sqrt{\int{\lambda~R~d\lambda}\over
\int{\lambda^{-1}~R~d\lambda}}~.\label{eq3}\end{equation} The conversion from
$F_{\nu}$ to $F_{\lambda}$ is \begin{equation}\langle
F_{\lambda}\rangle={c\over\lambda_p^{2}}\langle
F_{\nu}\rangle~,\label{eq4}\end{equation} where $\lambda_p \nu_p=c$ and c is the
speed of light.  In $cgs$ units with $\langle F_{\lambda}\rangle$ in
$(erg~cm^{-2}~s^{-1}~\AA^{-1})$, $\langle F_{\nu}\rangle$ in
$erg~cm^{-2}~s^{-1}~Hz^{-1}$, $c=2.99792~10^{18}~\AA~s^{-1}$, and 
$1~mJy=10^{-26}~erg~cm^{-2}~s^{-1}~Hz^{-1}$, the final conversion from mJy to
CALSPEC units is \begin{equation}\langle F_{\lambda}\rangle
(erg~cm^{-2}~s^{-1}~\AA^{-1})=
{2.99792~10^{-16}\over\lambda_p^{2}}\langle~F_{\nu}\rangle(mJy)~.\label{eq5}\end{equation}
Thus, the mean flux $\langle F_{\lambda}\rangle$ from a CALSPEC spectral energy
distribution (SED) and Equation 1 can be compared to the IRAC photometry from
Equation 5 with $\lambda_p$ in microns.

\section{Data Reduction}			

All IRAC observations of 42 CALSPEC stars were downloaded from the Spitzer
Heritage
Archive{\footnote{https://sha.ipac.caltech.edu/applications/Spitzer/SHA/}, where
these data are the flux files (*\_cbcd.fits and *\_bcd.fits), the data quality
files (*\_bimsk.fits), uncertainty files (*\_bunc.fits), and the mosaic files 
(*\_maic.fits), as described in the IRAC Instrument
Handbook{\footnote{https://irsa.ipac.caltech.edu/data/SPITZER/docs/irac/
iracinstrumenthandbook/} (IHB). Our stellar flux values are derived from the
*\_bcd.fits files, except for $\eta$ UMa, where the saturation corrected
\_cbcd.fits files are required. In general, the data reduction steps follow
those of \citet{krick2021}, including omission of the first frame of a series,
calculating the approximate positions from the mosaics or SIMBAD coordinates
with proper motions, refining the positions with a centroiding algorithm,
converting from the original units of MJy/sr to mJy by multiplying by 0.0235044
times the pixel area in arcsec, applying the array location-dependent
correction{\footnote{https://irsa.ipac.caltech.edu/data/SPITZER/docs/irac/calibrationfiles/
locationcolor/}, the pixel phase
correction{\footnote{https://irsa.ipac.caltech.edu/data/SPITZER/docs/irac/calibrationfiles/
pixelphase/}, an aperture correction, and a time-dependent correction for [3.6]
and [4.5]. The baseline time for a unity time-dependent correction is 2008.2,
which is the average date for the nine years of flight data used by
\citet{carey2012} to establish the absolute flux calibration. The aperture
photometry utilizes the \citet{krick2021} radius of three pixels, and
the individual photometry measures for each channel are combined using an
iterative $3\sigma$ clipped mean for the final result. A three pixel
photometry radius is chosen because of the heritage, because the IRAC
documentation provides the flux calibration, and because that radius is a good
compromise between enclosing the majority of the signal and avoiding
contamination from nearby stars.

The pixel phase correction for IRAC1 and IRAC2 is done with the code
irac\_aphot\_corr.pro{\footnote{https://irsa.ipac.caltech.edu/data/SPITZER/docs/dataanalysistools/
tools/contributed/irac/iracaphotcorr/}, while the four files
ch*\_photcorr\_rj.fits for the cold mission and the two files 
ch*\_al\_s192.fits{\footnote{https://irsa.ipac.caltech.edu/data/SPITZER/docs/irac/calibrationfiles/locationcolor}
for the warm mission define the location-dependent corrections.

An additional correction to warm mission IRAC1--2 subarrays is for improper
dark removal in the Archival products for staring mode observations
\citep{krick2021}. The standard dithered darks must be removed, and then the
staring darks are applied using the code
change\_dark\_calibrate.pro{\footnote{https://irsa.ipac.caltech.edu/data/SPITZER/docs/dataanalysistools/tools/contributed/irac/change\_dark\_calibrate/}.
Three stars, 1808347, BD+60$^{\circ}$1753, and HD165459, have staring
mode observations for warm mission subarrays; and applying the dark correction
makes a significant difference for only the one case of 1808347
IRAC1 with a reduction of 1.7\% in flux.

\subsection{Sky Background}

The measurement of the sky background is crucial for obtaining precise
photometry, especially for faint stars where this background reaches levels of
$\approx$100 times the sky-subtracted net signal within the three pixel
photometry radius. For example, the IRAC4 sky for G191B2B is 11 times the net
photometry of 0.44 mJy; and for the fainter star, LDS749B, the sky is 170 times
the net mJy in IRAC4, causing an 18\% uncertainty. To get the best estimates of
sky level, this background is computed for the sky annulus before doing the
three pixel radius photometry and then input into the photometry routine,
instead of using any black-box algorithm within the photometry code. While
the median of the pixel values in the designated sky annulus is often a
good estimator, our choice for the sky background is an iterative $3\sigma$
clipped mean with a robust uncertainty from the rms standard deviation of
the unclipped pixel values. The clipping is carefully checked to avoid rejection
of valid data.

\subsection{Aperture Corrections}

The absolute flux calibration of the Spitzer Archival IRAC products is from 
\citet{carey2012}, where 10-pixel radius photometry with a 12--20 pixel sky
annulus is the reference photometry for the Archival IRAC images. Following 
\citet{krick2021}, our adopted photometry radius is three pixels, which reduces
sky noise. For our fainter stars with a signal near the sky brightness
level, a radius as large as 10 pixels has an uncertainty that is dominated by
noise from the multitude of background pixels.

Table 4.8 of the IHB contains the aperture corrections from three to the 
standard 10 pixel radius with a choice of two background annuli of 3--7 or
12--20 pixels. The 12--20 background annulus is preferred because of the large
separation from any signal in the PSF wings, and because the larger number of
pixels increases the precision of our $3\sigma$ clipped mean background.
However, for the small 32x32 pixel subarray data, the larger background annulus
always extends beyond the boundaries of the image and necessitates a 3--7 pixel
sky region for subarrays.

This three pixel minimum radius for the sky is problematic because of the
residual signal in the PSF wings at a three pixel radius and because of the
details of how the partial pixels are computed for the sky. To derive the best
aperture correction for our photometry methods, the ratios of full-frame
(256x256 pixel) photometry for 3--7 pixel vs. 12--20 pixel sky annuli for 11
well-observed stars appear in Figure~\ref{sky7vs20}. The weighted mean values
are the horizontal dashed lines. For IRAC3, only one point for the faint 1743045
at X= 7.0e-05 lies more than $1\sigma$ from the mean; and only the even fainter
1743045 IRAC4 at X= 7.5e-05 lies more than $2\sigma$ from the mean. The four
mean values are written in each panel along with the uncertainty that is a
maximum of 0.003. These differences from unity are equivalent to adjustments of
the Table 4.8 of the IHB aperture correction for the 3--7 sky  annulus from
[1.125,1.120,1.135,1.221] to [1.146,1.140,1.157,1.248] for the four IRAC
channels in the cold mission. For the warm mission IRAC1--2, the corresponding
change is from [1.1233,1.1336] to [1.144,1.154]. These  $\approx2\%$ increases
in subarray fluxes have uncertainties of $<$0.3\% and are incorporated in our
results. The difference in values between this work and the IHB are likely
caused by subtle differences in the handling of partial pixels.

\begin{figure}			
\includegraphics*[width=.95\textwidth,trim=0 50 0 20]{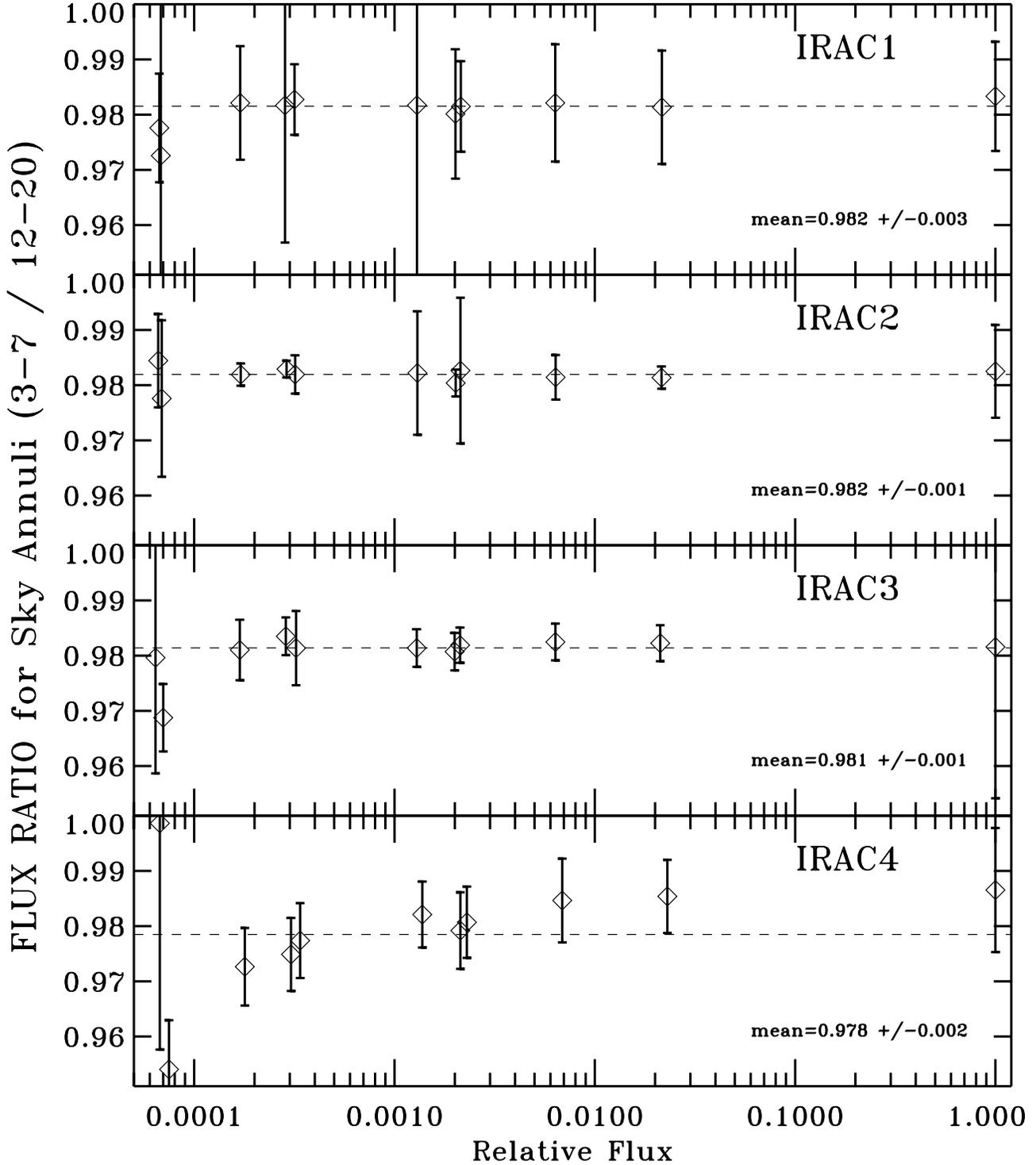}
\caption{\baselineskip=12pt
Ratio of three-pixel radius photometry for 3--7 pixel vs. 12--20 pixel sky
annuli in the four Spitzer IRAC bands. Rarely do points fall off the weighted
average dashed lines by more that the $1\sigma$ error bars, even for IRAC4,
where there is a hint of a small flux dependent trend. The Relative Flux
values are normalized to the photometry for the brightest star $\eta$ UMa
at X=1.000.
\label{sky7vs20}} \end{figure}

\subsection{Stellar Centroids}		

For all but the faintest stars with an in-band flux of
$<2\times10^{-18}~erg~cm^{-2}~s^{-1}~\AA^{-1}$ erg, a signal-weighted,
center-of-light average in x and y locates the stellar centers and provides
similar results to the standard, but somewhat balky IRAC-specific
box\_centroider.pro, which sometimes finds positions outside of the image. For
the faintest observations, where centroid positions are dominated by noise,
forced photometry is required . The astrometric position of the star that
includes the proper motion to the epoch of the observation is used as the
aperture centroid position. From the CALSPEC SED, a division at
$2\times10^{-18}~erg~cm^{-2}~s^{-1}~\AA^{-1}$ in each band separates the use of
found centroids from forced photometry. For IRAC1, this limit corresponds to
0.085 mJy, but all  of our stars are brighter, so centroiding always applies for
IRAC1.

Stellar positions derived independently by the first three authors from the
mosaic images are compared to the astrometric positions. A typical scatter of
$\approx0.2\arcsec$ exists among both the right ascension (RA) and declination
(DEC) of the three different results, even when the mosaic position is found
with similar software. Consequently, an  uncertainty of 0.2\arcsec\ is adopted
in each axis; and photometry at the center and four corners of a box offset by
$\pm0.2\arcsec$ in RA and DEC determines an uncertainty corresponding to the
range of the photometry at the five points. Because the error in measuring
centroid position is sometimes larger than the typical $0.2\arcsec$, our added
uncertainty is a minimal estimate. This uncertainty in the stellar position is
combined in quatrature with the
$(rms~scatter~of~the~flux~values)/sqrt(number~of~obs)$ to get a total error bar.
Because the phase correction is more strongly peaked for IRAC1 than for IRAC2,
the uncertainties that are ascribed to shifts in IRAC1 position are systematically
larger than for IRAC2.

\subsection{Time-dependent Correction}		

\citet{krick2016,krick2021} recommend a time-dependent correction for IRAC1
[3.6] and IRAC2 [4.5] of -0.1 and -0.05 \%/year for the loss of sensitivity over
the time period covering both the cold and warm missions. IRAC3--4 do not show
significant degradation over the shorter time period of the cold mission. To
refine the loss rates, Figure~\ref{aorflx1} for IRAC1 and
Figure~\ref{aorflx2} for IRAC2 show the change of response vs. time for four
stars over the 2004--2020 time frame, where the photometry includes the
\citet{krick2021} loss rates of -0.1 and -0.05\% /year for IRAC1 and IRAC2. Our
observed fluxes binned by Astronomical Observation Request (AOR) to increase the
S/N are normalized by dividing by the predicted flux of
the CALSPEC SEDs.

\begin{figure}			
\includegraphics*[width=.95\textwidth,trim=0 90 0 20]{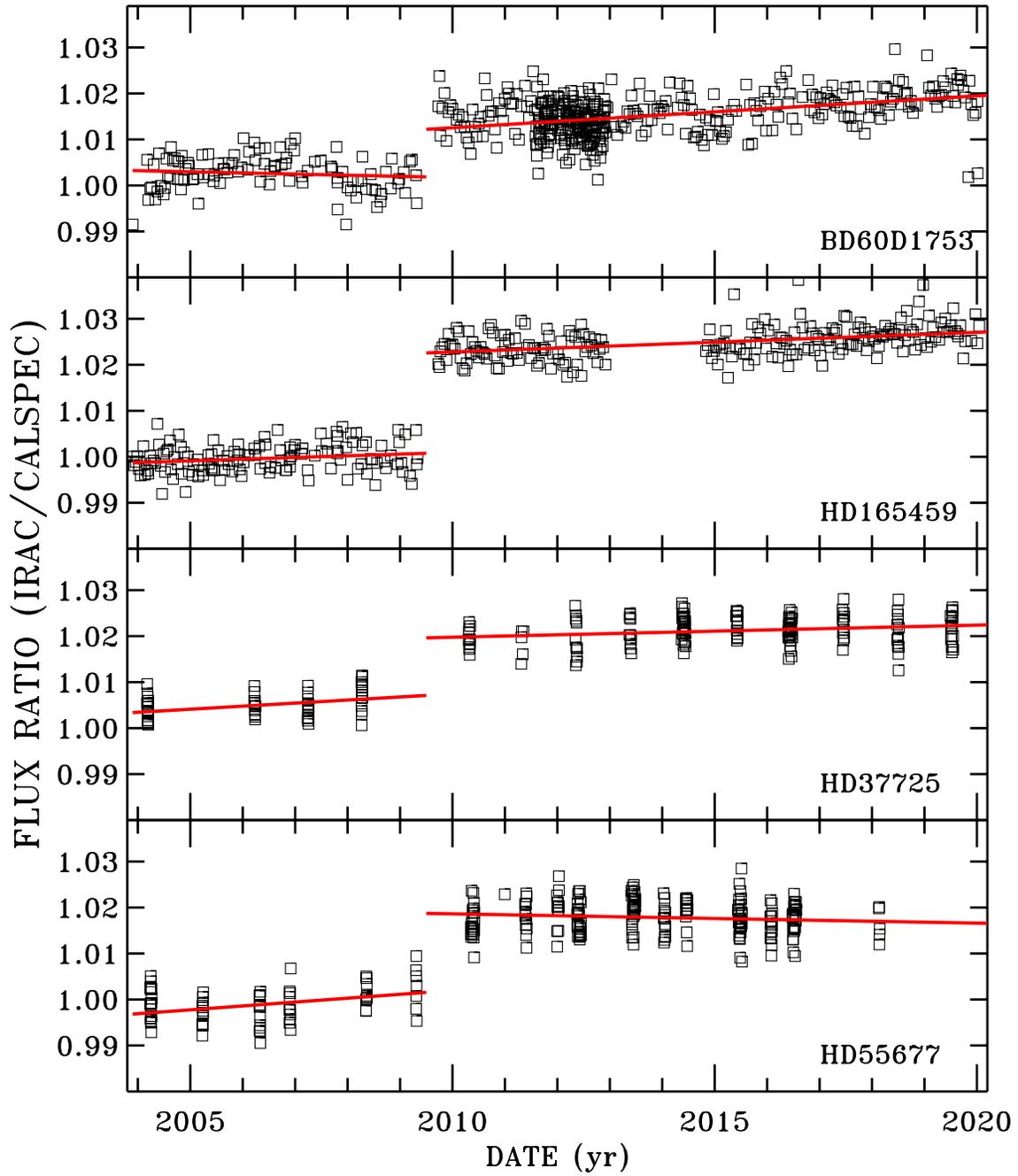}
\caption{\baselineskip=12pt
Ratio of our IRAC1 three-pixel radius photometry to the predicted CALSPEC flux
for four well observed stars with exposure times of 1.2s, except for HD165459
with 0.2s exposures. This full-frame photometry is binned by AOR to increase the
S/N. The red lines are separate linear fits for the cold and warm missions. 
Only the standard recommended IRAC calibration parameters are included in 
converting the raw photometry to flux values.
\label{aorflx1}} \end{figure}

\begin{figure}			
\includegraphics*[width=.95\textwidth,trim=0 90 0 20]{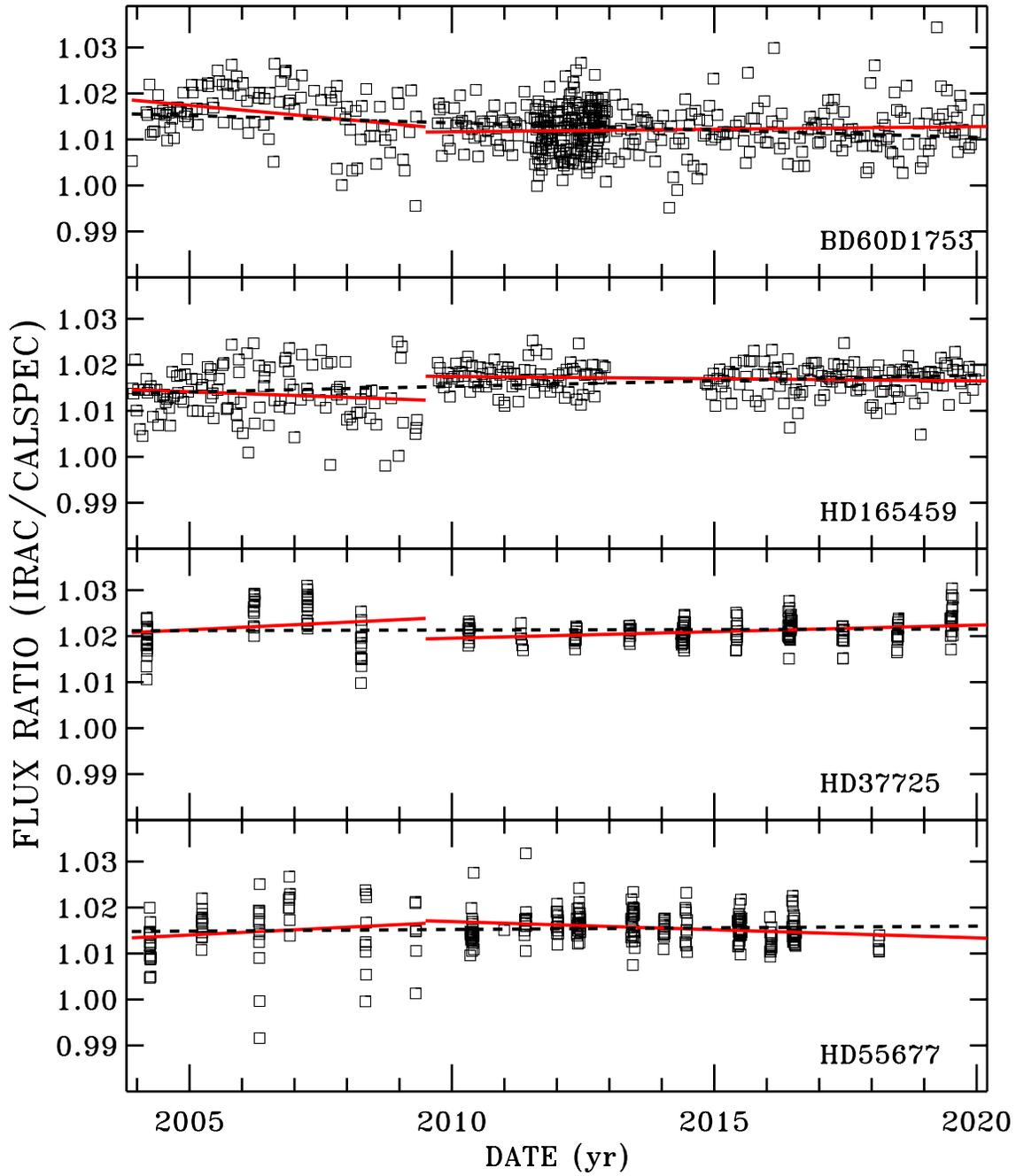}
\caption{\baselineskip=12pt
Ratio of our IRAC2 three-pixel radius photometry to the predicted CALSPEC flux
for four well observed stars, as in Figure~\ref{aorflx1}. The photometry is
binned by AOR to increase the S/N. Only the standard recommended IRAC calibration
parameters are included in  converting the raw photometry to flux values.
\label{aorflx2}} \end{figure}

\begin{figure}			
\includegraphics*[width=.95\textwidth,trim=0 90 0 20]{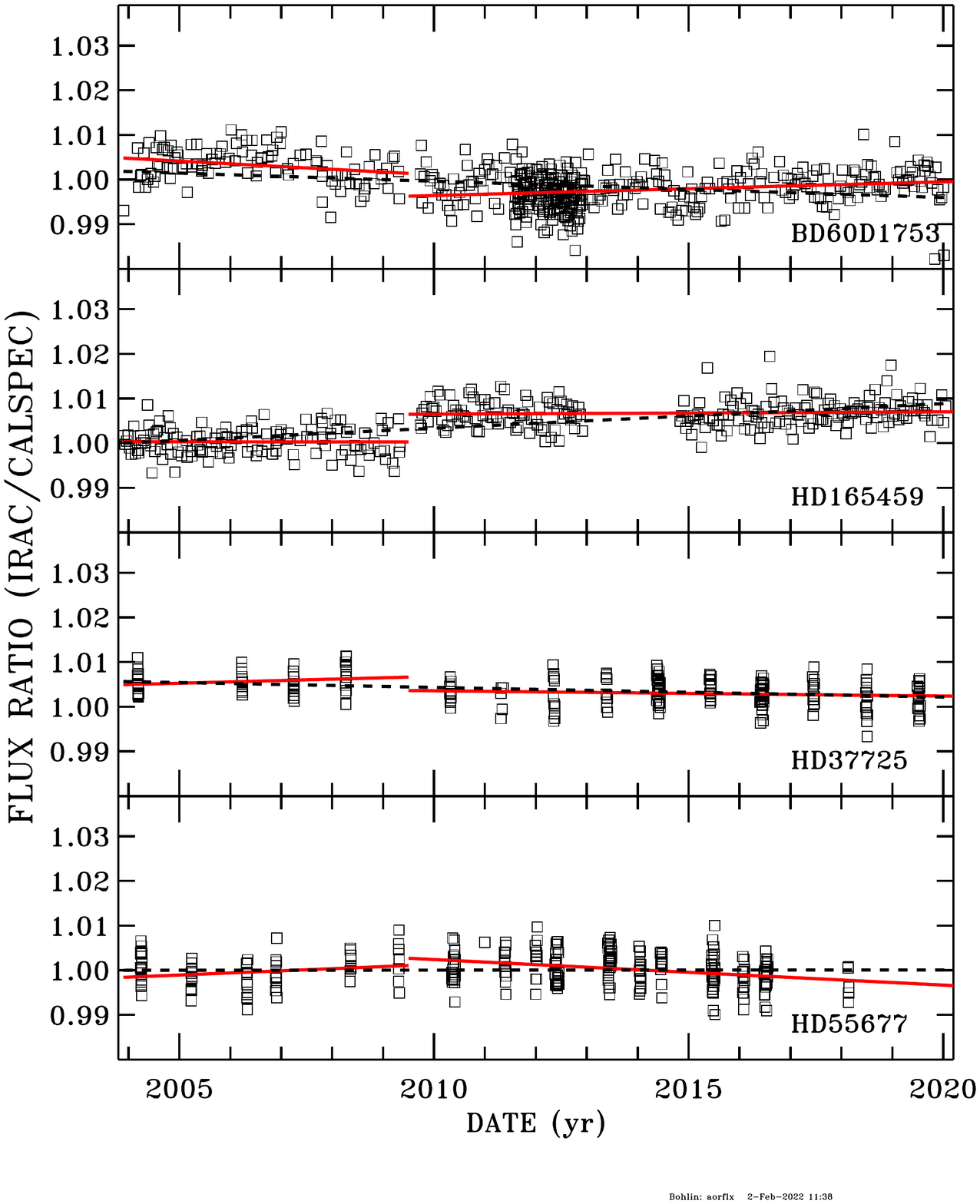}
\caption{\baselineskip=12pt
As in Figure~\ref{aorflx1} after correcting for the discontinuity jump of 1.55\% 
at 2009.5 and for the new IRAC1 time-dependent loss of sensitivity of 0.063
\%/yr. The pair of red line fits match the single dashed line fit to $<$0.4\%;
and the largest deviation from a constant for the dashed lines over the entire
16 years approaches 1\% only for HD165459 near 2020.
\label{aorflx1fix}} \end{figure}

Table~\ref{table:tcor} includes the statistics of our red-line linear fits in 
Figures~\ref{aorflx1}--\ref{aorflx2}, where the slope columns are in \%/year.
The cold and warm time periods are fit and tabulated separately as slope1 and
slope2 for IRAC1--2 in Table~\ref{table:tcor}, while the columns labeled as Jump
measure the amount of discontinuity between the fits at 2009.5. The average and
uncertainties for the discontinuous jumps at 2009.5 in
Figures~\ref{aorflx1}--\ref{aorflx2} are 1.55 (0.25) and -0.39 (0.20)\% for
IRAC1--2 with the uncertainties in parentheses. These adjustments are
implemented as updates to the warm mission fluxes after 2009.5. 

After correcting for the discontinuous jumps and fitting over the complete time
period, the average slopes and 1$\sigma$ uncertainties are +0.037 (.020) and
-0.010 (.012), for IRAC1--2. Thus, the  refined IRAC1 slope value becomes
-0.10+0.037=-0.063, while the Krick value of -0.05 for IRAC2 becomes
-0.05-0.010=-0.060 \%/year. Neither new loss rate differs by more than
2$\sigma$ from the Krick value.

Figure~\ref{aorflx1fix} is an example of the results after implementing the new
corrections for discontinuities and for the newly derived rates of sensitivity
loss. In the worst case of the 0.2s exposures of HD165459, the range of IRAC1
photometry as traced by the dashed line in Figure~\ref{aorflx1fix} is 1\%, which
represents a reduction from the 3\% spread in Figure~\ref{aorflx1} before
correction. Section 6.3 discusses the validity of the existing IRAC absolute
flux calibration with our adjustments included. The 1\% deviation of the dashed
line fit from a constant for HD165459 over the whole 2004--2020 period
represents an estimate of the systematic uncertainties of our results.

\newpage
\subsection{Repeatability}		

The rms scatter about the dash-line fits in Figure~\ref{aorflx1fix} provide a
measure of IRAC repeatability for observations at different epochs and different
locations on the detector after making our corrections to the Archival IRAC
fluxes. These rms measures range from 0.3--0.4\% for IRAC1 and 0.4--0.5\% for
IRAC2. Thus, 0.3--0.5\% reduced by sqrt(N-independent observations binned by
AOR) could be included in estimates of the IRAC photometric uncertainty whenever
repeatability is a concern. Because these repeatability estimates include some
noise sources, repeatability should be improved for cases with higher S/N.
However, sub-percent repeatability is already impressive and sets an excellent
limit on how precise IRAC results can be with  proper photometric corrections
and sufficient S/N.

\begin{deluxetable}{lcccccccc}		
\tabletypesize{\normalsize}
\tablewidth{0pt}
\tablecolumns{9}
\tablecaption{\label{table:tcor} Sensitivity Changes}
\tablehead{
\colhead{Star} & \colhead{Exp (s)} & 
\multicolumn{3}{c}{3.6 \micron} &
\multicolumn{1}{c}{    } &\multicolumn{3}{c}{4.5 \micron} \\
     &  &\colhead{Slope1} &\colhead{Slope2} &\colhead{Jump(\%)} &\colhead{    }
	&\colhead{Slope1} &\colhead{Slope2} &\colhead{Jump(\%)}}
\startdata
BD+60$^{\circ}$1753 &1.2 &-0.024 &+0.069 &1.03 &      &-0.114 &+0.001 &-0.506 \\
HD165459            &0.2 &+0.036 &+0.043 &2.18 &      &-0.051 &-0.020 &+0.126 \\
HD37725             &1.2 &+0.067 &+0.026 &1.25 &      &+0.044 &+0.018 &-0.839 \\
HD55677             &1.2 &+0.084 &-0.020 &1.72 &      &+0.047 &-0.045 &-0.340 \\
\enddata
\end{deluxetable}

\subsection{Exposure Time Corrections}			

Both the IHB and \citet{krick2021b} point out differences in photometry that
depend on exposure time and on subarray vs. full-frame data for channels 1 and
2. Here, a more detailed analysis reveals the photometry dependence on
individual exposure times in both subarray and full-frame modes for all four
IRAC channels. For the few stars in our sample with robust sets of observations
at more than one exposure time, the average flux for each AOR is divided by the
average flux at the maximum exposure time for that particular star and channel
to form a set of relative flux values as plotted in Figure~\ref{expcor} as small
black diamonds. For stars where the maximum exposure time is 0.08 s, those pairs
of relative flux at 0.01 and 0.08 s are normalized to the average of all other
stars at the exposure level of 0.08 s. The large red diamonds at each exposure
time are the averages over all stars for which there are data, where the values
of unity at the normalization levels of the maximum exposure times are omitted
from the averages.

\begin{figure}			
\includegraphics*[width=.95\textwidth,trim=0 60 0 20]{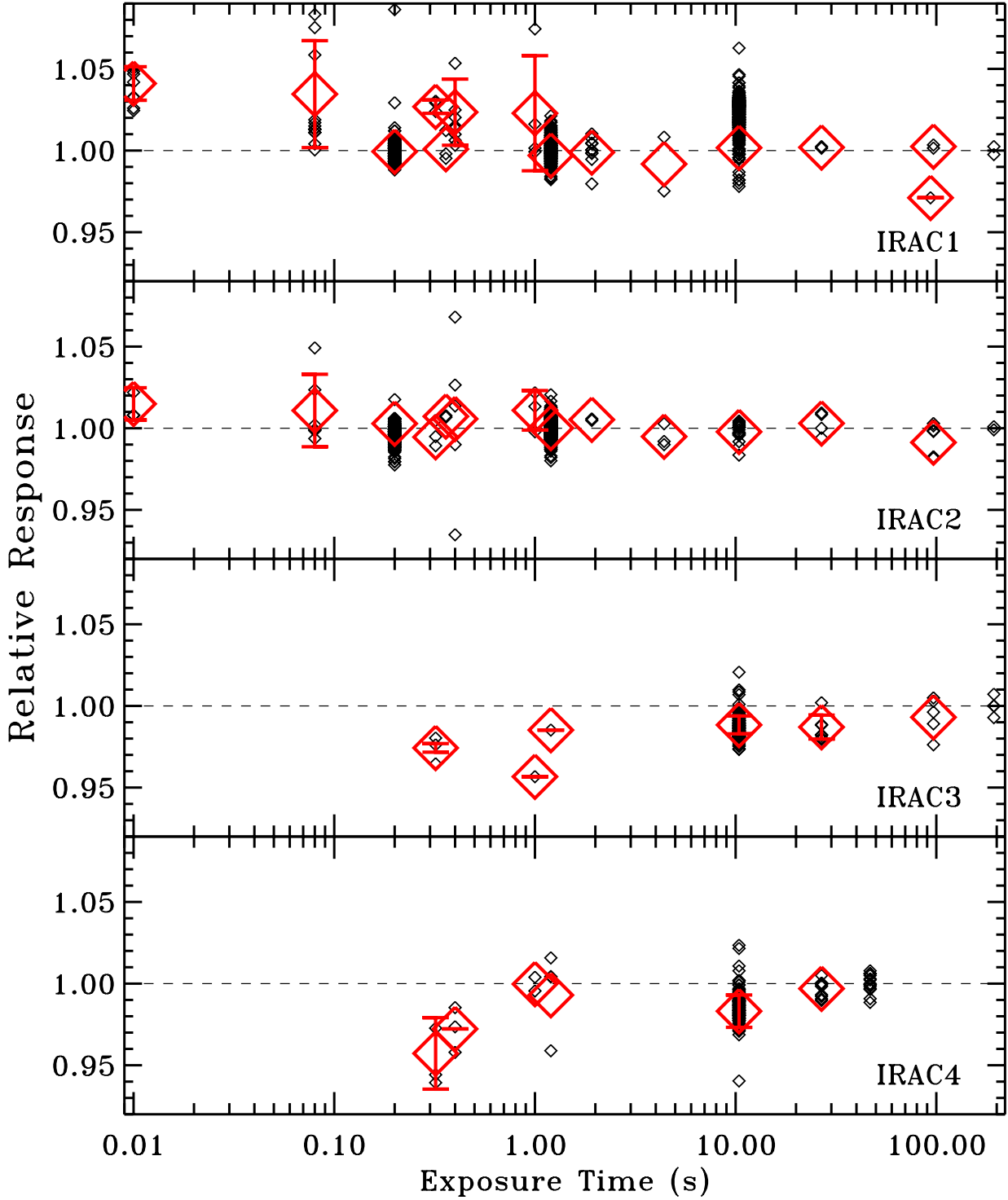}
\caption{\baselineskip=12pt
Ratio of our IRAC photometric fluxes to the average IRAC flux for the longest
exposure for each star (small black diamonds) for the few stars with robust sets
of observations at two or more exposure times. The photometry is binned by AOR
to increase the S/N. The large red diamonds are the ratios averaged over the
Nstar stars at each measured exposure time from Table~\ref{table:expcor}. The
error bars are the $1\sigma$ rms scatter among all the contributing stars and
appear for corrections of $\ge1\%$. The results that are based on only one star
have error bars with zero extent. Occasionally, the red diamonds are offset from
the  centroids of the black points, because the average of all stars is offset
from the average of a star with a dominant number of observations.
\label{expcor}} \end{figure}

\begin{deluxetable}{rcccr}		
\tabletypesize{\scriptsize}
\tablewidth{0pt}	
\tablecolumns{5}	
\tablecaption{\label{table:expcor} Exposure Time Adjustments}
\tablehead{
\colhead{Exposure (s)} & \colhead{Adjustment} & 
     \colhead{rms} &\colhead{Nstar} &\colhead{Ntot} }

\startdata
IRAC1  & & & & \\
    0.01 &  \textbf{1.041} &   0.009 &       4 &    2318 \\
    0.08 &  \textbf{1.035} &   0.032 &       4 &   13069 \\
    0.20 &   0.999       &   0.004 &       1 &    4988 \\ 
    0.32 &  \textbf{1.027} &   0.004 &       2 &    1131 \\
    0.36 &   1.001       &   0.004 &       2 &    2071 \\
    0.40 &  \textbf{1.024} &   0.020 &       4 &      40 \\	  
    1.00 &   1.023       &   0.035 &       4 &      16 \\	
    1.20 &   0.997       &   0.002 &       4 &    4551 \\
    1.92 &   0.999       &   0.006 &       2 &    2386 \\
    4.40 &   0.992       &   0.023 &       1 &     685 \\	
   10.40 &   1.002       &   0.014 &       4 &    2842 \\
   26.80 &   1.002       &   0.001 &       1 &      58 \\	
   93.60 &  \textbf{0.971} &   0.004 &       1 &     130 \\	
   96.80 &   1.002       &   0.001 &       1 &      36 \\	
  193.60 &   1.000       &   0.004 &       1 &      20 \\	
IRAC2 & & & & \\
    0.01 &  \textbf{1.015} &   0.010 &       2 &    1068 \\
    0.08 &   1.011       &   0.022 &       3 &    9983 \\ 
    0.20 &   1.003       &   0.008 &       3 &    5050 \\
    0.32 &   0.995       &   0.007 &       2 &    1132 \\
    0.36 &   1.007       &   0.001 &       1 &    1505 \\	
    0.40 &   1.006       &   0.006 &       2 &      20 \\	  
    1.00 &   1.011       &   0.012 &       3 &      12 \\	
    1.20 &   1.000       &   0.004 &       5 &    4735 \\
    1.92 &   1.005       &   0.005 &       1 &     251 \\	
    4.40 &   0.995       &   0.023 &       1 &     139 \\	
   10.40 &   0.998       &   0.005 &       3 &     294 \\
   26.80 &   1.003       &   0.008 &       2 &      88 \\
   96.80 &   0.991       &   0.011 &       1 &     165 \\	
  193.60 &   1.000       &   0.001 &       1 &      20 \\	
 IRAC3 & & & & \\
    0.32 &  \textbf{0.967} &   0.013 &       2 &    1132 \\
    1.00 &  \textbf{0.944} &   0.051 &       1 &       4 \\ 
    1.20 &   0.972       &   0.033 &       1 &       4 \\	
   10.40 &  \textbf{0.985} &   0.010 &       4 &     725 \\
   26.80 &  \textbf{0.987} &   0.007 &       3 &     108 \\
   96.80 &   0.993       &   0.010 &       3 &      55 \\
  193.60 &   1.000       &   0.018 &       3 &      25 \\ 
IRAC4 & & & & \\
    0.32 &  \textbf{0.958} &   0.023 &       2 &    1133 \\
    0.40 &  \textbf{0.973} &   0.014 &       1 &      12 \\
    1.00 &   1.000       &   0.006 &       2 &       8 \\	
    1.20 &   0.987       &   0.028 &       3 &     339 \\
   10.40 &  \textbf{0.983} &   0.010 &       4 &     714 \\
   26.80 &   0.997       &   0.004 &       4 &     118 \\
   46.80 &   1.000       &   0.001 &       4 &     298 \\
\enddata
\tablecomments{\textbf{Bold} adjustments to derived fluxes are implemented in
our data analysis.} 
\end{deluxetable}

Table~\ref{table:expcor} summarizes the results of the exposure level analysis,
where the 13 values in \textbf{bold} represent the corrections that are
implemented in the production of final flux values. The adjustment amounts are
those $\ge$1\% with rms values of less than the adjustment. The correction is in
the denominator of the flux calculation. For the 0.08 s exposure time, i.e. 0.1
s frame time, our results are in accord with the elevated normalized flux in
figure 3 of \citet{krick2021b} for IRAC1--2. The other columns in
Table~\ref{table:expcor} are for the \textit{rms} scatter among the
\textit{Nstar} count of stars with data at that exposure time. For exposure
levels with \textit{Nstar}=1, the rms value is the scatter among the individual
AOR measures at that level. \textit{Ntot} is the total number of useful
individual observations in all of the available AORs, where each subarray data
set has 64 individual observations. While most of the required adjustments for
IRAC1--2 are for the shorter subarray exposures, an exception is the 93.6 s time
for IRAC1 at 0.971. In the top panel of Figure~\ref{expcor}, the red diamond for
the 93.6 s exposure time encompasses the single black square AOR. However, this
single AOR includes 130 separate observations of 1743045 and is robust with an
rms of only 0.4\% among the 130 observations. After making this 3\% correction,
the average flux for the 93.6 s observations agrees to $<$1\% with the average
IRAC1 fluxes for 1743045 at five other exposure levels that range from
1.2--193.6 s.

Even though some of our important IRAC 3--4 data have only 0.01 s exposures, our
data collection unfortunately includes no observations to calibrate the IRAC3--4
0.01 s exposure level relative to longer times. These average fluxes are
arbitrarily assigned an uncertainty of 3\%, a typical value for short
exposures. 

\subsection{Saturation}			

Bright stars with excessive exposure times saturate the IRAC detectors and may
produce bad photometry. The IRAC saturation level is defined as 90\% of the
full-well depth of the detectors, where the full-well depths in electrons for
IRAC1--4 from  Section 2.3.1 of the  IHB are 145000 (110000), 140000 (125000),
170000, and 200000 electrons, respectively, with the warm mission in
parentheses. Combining the 90\% full-well depths with the sensitivities 770
(840), 890 (730), 420, and 910 (electrons/sec)/(mJy) from the IHB Table 2.4 and
the percent signal in the peak pixel of 42 (37), 43 (34), 29, and 22 from the
IHB Table 2.1 provides the IRAC1--4 saturation limits in (mJy s) of 404 (319),
329 (453), 1256, and  899. Dividing these (mJy s) limits by the predicted flux
from the CALSPEC SED in mJy from Equations 1 and 5, yields the exposure time to
saturation for each band.

 
An alternative, set of (mJy s) limits accrue from multiplying the exposure time
by the IHB Table 2.11 tabulated maximum unsaturated point source flux in mJy, as
calculated from the IHB equation 2.14. For example for IRAC1, these alternative
limits range from 190--368 (192--288) in comparison to the above 404 (319). Even
though puzzling to the neophyte, the IHB suggests that the saturation limits
depend on frame time, instead of exposure time; and saturation limits in frame
time multiplied by tabulated maximum unsaturated point source flux in mJy are
much closer to our 404 (319), ie. 379--400 (280--460). There is still a lot of
scatter but the values agree much better with our estimates. Another example
compares our IRAC4 (mJy s) limit of $\approx$900 to the IHB range of 450--5420
for exposure time or 900--5600 for frame time. Because the IHB includes the
effects of the background current that contributes to the filling of the
detector well capacity, the saturation limits should be less, NOT more than our
900 that ignores the background contribution. Because of the ambiguities
associated with Table 2.11 plus the omission of some important exposure times,
like 1.0 and 46.8 s, our simple-minded estimates define our saturation levels.
Any adjustments to our values would just affect the rare inclusion or exclusion 
of small bits of data in the measured IRAC fluxes.

The 46,122 IRAC1 and 44,145 IRAC2 observations for HD165459 cover a wide range
from under- to overexposure. For the predicted IRAC1--2 fluxes of 643, 416 mJy,
the saturation limits to 90\% full-well depth of the peak pixel are IRAC1 0.63s
(0.50s) and IRAC2 0.79s (1.09s). Table~\ref{table:hd165} illustrates the measured
fluxes from minimum to the maximum exposure time, where a blank row separates
the saturated from the unsaturated data. For the saturated data, a correction
for saturation is required, so the saturated flux measures utilize the
\_cbcd.fits files. The unsaturated flux measures in Table~\ref{table:hd165} and
the robust values for the longest exposures in the last row all agree to 2\% for
both IRAC1 and IRAC2. These long exposures of 23.6 and 26.8 s have overexposure
factors of 37 (47) and 34 (25) for each channel with the warm mission in
parentheses.

All the measured photometric flux values in Table~\ref{table:irac_fluxes}
utilize only unsaturated IRAC data, except for $\eta$ UMa, which has only
saturated data. The overexposure levels for the existing 10.4s exposures for all
channels 1--4 have overexposure factors between 780 and 77 and enable accurate
photometry for $\eta$ UMa from the *cbcd.fits files. However, an additional
uncertainty of 2\% is included for the $\eta$ UMa fluxes because of possible
systematic errors in the use of saturated data. 

In Table~\ref{table:irac_fluxes}, Unc is the 1$\sigma$ uncertainty, N is the
total number of observations used to find the average Flux, and Ratio is IRAC
photometry divided by CALSPEC synthetic photometry. Except for the few arbitrary
estimates of systematic uncertainty, Usys, assigned for variability, saturation
in $\eta$ UMa, or lack of correction for the IRAC3--4 short exposure times, the
uncertainties are just statistical. The statistical contributions arise from
Uobs, the rms scatter among the observations used to measure the flux as divided
by the square root of the number observations, N, and from Uoff, the total range
of the set of five photometry values with the 0.2\arcsec\ offsets. The final
uncertainty estimates are $sqrt(Uobs^2+Uoff^2+Usys^2)$. In general, Uobs is
negligible in comparison to Uoff, except for the cases with less than
$\approx$10 observations, while most cases have Usys=0. Many of these error
estimates seem too low, possibly because of unidentified systematic errors, such
as errors in the IRAC system fractional throughput and such as possible
non-linearity for weak exposures, where only a small fraction of the detector
full-well capacity is filled.

IRAC exposures with exposure times less than a factor of 14 below the maximum
unsaturated exposure contribute only noise and are not included in our analyses.
In order to include the 2257 faint 0.01 s exposures and augment the three
unsaturated 0.20s exposures, the acceptance level is

\begin{deluxetable}{ccccrccccrc}	  
\tabletypesize{\normalsize}
\tablewidth{0pt}	
\tablecolumns{11}
\tablecaption{\label{table:hd165} HD165459 Flux vs. Exposure Time}
\tablehead{\colhead{Frame time (s)} &
\colhead{Exptim (s)} & \multicolumn{4}{c}{3.6 \micron} &
\multicolumn{1}{c}{    } &\multicolumn{4}{c}{4.5 \micron} \\ &
  \colhead{  } &   \colhead{Flux (mJy)} & \colhead{Unc (\%)} & \colhead{N} & 
\colhead{Ratio}& \colhead{        } & 
\colhead{Flux (mJy)} &  \colhead{Unc (\%)}& 
\colhead{N} & \colhead{Ratio}}
\startdata
0.02   &0.01 &644.0 &0.63 &1316& 0.996 &     &421.0 &0.37 &817   & 0.993 \\
0.10   &0.08 &631.8 &2.27 &4300& 0.977 &     &423.8 &0.66 &2381  & 1.000 \\
0.40   &0.20 &646.1 &0.73 &3500& 0.999 &     &422.4 &0.12 &3394  & 0.997 \\
0.40   &0.32 &644.6 &0.23 &252 & 0.997 &     &419.2 &0.20 &252   & 0.989 \\
0.40   &0.36 &646.5 &1.15 &32423& 1.000&     &423.7 &0.28 &32908 & 1.000 \\
   &  &  &  &  &  &  &  &  &  &  \\
2   &1.20 &520.8 &4.54 &8  & 0.806 &      &415.9 &0.12 &601   & 0.982 \\
6   &4.40 &697.0 &2.32 &9  & 1.078 &      &428.2 &1.50 &9     & 1.011 \\
12   &10.4 &681.3 &2.99 &9  & 1.054 &      &431.7 &1.02 &9     & 1.019 \\
30   &23.6,26.8\tablenotemark{a} &656.7 &0.41 &98  & 1.016&     &427.6 &0.36 &101 &1.009\\
\enddata
\tablenotetext{a}{26.8 is the Exposure time for IRAC2 4.5 \micron}
\tablecomments{Unc is the 1$\sigma$ uncertainty, N is the number of observations
used, and Ratio is the flux normalized to the flux for the 0.36 s exposures.}
\end{deluxetable}

relaxed from a factor of 14 to 20 below 0.20s for HD14943 IRAC2.

\begin{deluxetable}{lrrrrrrrrrrrrrrrr}		
\tabletypesize{\scriptsize}
\tablecaption{Measured IRAC Fluxes \label{table:irac_fluxes}}
\tablehead{\colhead{Star} & \multicolumn{4}{c}{3.6  \micron} &
\multicolumn{4}{c}{4.5  \micron} &  \multicolumn{4}{c}{5.8  \micron} &
\multicolumn{4}{c}{8.0  \micron} \\ &  
\colhead{Flux} & \colhead{Unc} & \colhead{N}  & \colhead{Ratio} & 
\colhead{Flux} & \colhead{Unc} & \colhead{N}  & \colhead{Ratio} & 
\colhead{Flux} & \colhead{Unc} & \colhead{N}  & \colhead{Ratio} & 
\colhead{Flux} & \colhead{Unc} & \colhead{N}  & \colhead{Ratio} \\&
\colhead{mJy} & \colhead{\%} & \colhead{}& \colhead{} &
\colhead{mJy} & \colhead{\%} & \colhead{}& \colhead{} &
\colhead{mJy} & \colhead{\%} & \colhead{}& \colhead{} &
\colhead{mJy} & \colhead{\%} & \colhead{}& \colhead{}}
\startdata
&HOT Stars & \\
     10LAC & 1.578e+03&  2.71&     63&  0.997&  1.010e+03&  1.32&     63&    1.011&  6.165e+02&  3.29&      63&    0.991&  3.276e+02&  3.22&     63&   0.972 \\
    ETAUMA & 2.955e+04&  2.22&     10&  0.975&  1.934e+04&  2.09&     10&    1.000&  1.267e+04&  2.80&      10&    1.039&  6.588e+03&  2.26&     10&   0.983 \\
   G191B2B & 1.999e+00&  0.61&    113&  0.992&  1.283e+00&  0.57&     63&    1.005&  8.081e-01&  1.17&      58&    1.009&  4.351e-01&  2.85&     37&   0.986 \\
     GD153 & 4.864e-01&  0.68&    108&  0.986&  3.097e-01&  0.50&     82&    0.993&  1.937e-01&  4.91&      59&    0.992&  1.082e-01&  7.23&     62&   1.009 \\
      GD71 & 6.714e-01&  0.73&     60&  0.994&  4.355e-01&  0.66&     69&    1.026&  2.763e-01&  3.27&      59&    1.042&  1.608e-01&  5.22&     64&   1.112 \\
    LAMLEP & 2.431e+03&  0.69&   2325&  0.995&  1.567e+03&  0.20&   2320&    1.014&  9.481e+02&  3.14&      63&    0.980&  5.148e+02&  3.29&     63&   0.975 \\
     MUCOL & 1.003e+03&  1.04&   2202&  0.993&  6.565e+02&  0.19&   2263&    1.031&           &      &	      &         &           &      &       &         \\
&A Stars  & \\ 
    109VIR & 9.297e+03&  0.83&   7413&  0.998&  6.219e+03&  0.21&   2506&    1.033&           &      &	      &         &	    &      &       &         \\
   1732526*& 3.486e+00&  1.66&     51&  0.975&           &      &       &         &           &      &	      &         &	    &      &       &         \\
   1743045 & 2.066e+00&  2.60&    244&  1.020&  1.343e+00&  1.16&    247&    1.022&  8.757e-01&  0.48&     116&    1.043&  4.908e-01&  0.63&    370&   1.052 \\
   1757132 & 9.436e+00&  0.44&    155&  1.014&  6.216e+00&  0.21&     95&    1.030&  4.023e+00&  0.36&      35&    1.044&  2.214e+00&  0.50&    107&   1.036 \\
   1802271 & 5.053e+00&  0.71&    101&  1.007&  3.300e+00&  0.15&    112&    1.017&  2.114e+00&  0.41&      48&    1.020&  1.173e+00&  0.49&    158&   1.021 \\
   1805292 & 4.398e+00&  0.84&     52&  1.002&  2.941e+00&  0.47&     33&    1.035&           &      &	      &	        &	    &      &       &         \\
   1808347*& 6.524e+00&  3.40&   1782&  0.997&  4.339e+00&  1.51&     23&    1.023&           &      &	      &	        &	    &      &       &         \\
   1812095*& 8.496e+00&  2.36&   3171&  1.002&  5.643e+00&  1.50&   2293&    1.027&  3.624e+00&  1.52&     590&    1.033&  1.990e+00&  1.57&    482&   1.024 \\
 BD60D1753 & 3.869e+01&  2.12&   5626&  0.992&  2.548e+01&  0.81&   4705&    1.011&  1.636e+01&  0.26&     519&    1.018&  9.103e+00&  0.44&    528&   1.023 \\
    DELUMI & 5.416e+03&  1.07&    252&  0.981&  3.598e+03&  0.48&    250&    1.007&           &      &        &         &           &      &       &         \\
   ETA1DOR & 1.362e+03&  0.57&    743&  0.992&  8.891e+02&  0.25&    503&    1.004&  5.665e+02&  0.20&     252&    1.005&  3.188e+02&  0.57&    250&   1.023 \\
  HD101452 & 5.180e+02&  1.20&      5&  1.021&  3.391e+02&  0.65&      4&    1.031&  2.158e+02&  0.36&       5&    1.027&  1.212e+02&  0.68&      4&   1.042 \\
  HD116405 & 1.122e+02&  0.60&     10&  0.996&  7.330e+01&  0.59&      8&    1.011&  4.652e+01&  0.41&       5&    1.010&  2.601e+01&  0.59&      4&   1.022 \\
  HD128998 & 1.303e+03&  0.34&   1045&  0.987&  8.690e+02&  0.55&    756&    1.019&  5.460e+02&  0.26&     566&    1.005&  3.085e+02&  0.56&    566&   1.025 \\
   HD14943 & 1.799e+03&  0.79&   2264&  1.001&  1.191e+03&  0.08&   2259&    1.024&  7.597e+02&  0.27&       3&    1.023&  4.271e+02&  0.72&      3&   1.038 \\
  HD158485 & 9.702e+02&  0.50&    484&  0.999&  6.458e+02&  0.39&    252&    1.027&  4.057e+02&  0.21&     252&    1.011&  2.307e+02&  0.55&    251&   1.037 \\
  HD163466 & 7.949e+02&  0.33&    862&  1.002&  5.307e+02&  0.18&    848&    1.033&  3.377e+02&  0.22&     567&    1.028&  1.895e+02&  0.82&    564&   1.040 \\
  HD165459 & 6.461e+02&  1.09&  38607&  1.004&  4.236e+02&  0.27&  38752&    1.019&  2.650e+02&  0.24&   11080&    0.999&  1.507e+02&  0.46&   1125&   1.026 \\
  HD180609 & 6.382e+01&  0.59&    792&  1.007&  4.176e+01&  0.99&   1407&    1.017&  2.693e+01&  0.24&     110&    1.027&  1.528e+01&  0.46&    522&   1.051 \\
    HD2811 & 4.216e+02&  1.34&      5&  1.011&  2.751e+02&  0.64&      4&    1.018&  1.760e+02&  0.53&       5&    1.020&  9.916e+01&  0.88&      4&   1.035 \\
   HD37725 & 1.895e+02&  0.75&   2060&  1.004&  1.248e+02&  0.28&   1856&    1.021&  8.053e+01&  0.24&     266&    1.032&  4.572e+01&  0.53&    244&   1.059 \\
   HD55677 & 6.005e+01&  0.83&   2692&  1.000&  3.954e+01&  0.17&   2447&    1.015&  2.528e+01&  0.25&     305&    1.016&  1.418e+01&  0.48&    298&   1.028 \\
&G Stars &\\
    16CYGB & 3.716e+03&  2.52&     63&  0.952&           &      &       &	  &	      &      &	      &         &  8.981e+02&  0.45&	  4&   0.995 \\
     18SCO & 7.032e+03&  1.22&    251&  0.929&           &      &	&	  &	      &      &	      &         &	    &      &	   &         \\
  HD106252 & 1.168e+03&  1.68&   4773&  0.978&  7.392e+02&  0.67&    250&    0.967&           &      &	      &         &  2.635e+02&  3.14&    251&   0.955 \\
  HD159222 & 2.633e+03&  1.13&   4768&  0.961&  1.697e+03&  0.14&    251&    0.972&           &      &	      &         &  5.973e+02&  3.26&    251&   0.944 \\
  HD205905 & 2.101e+03&  0.47&    250&  0.955&  1.353e+03&  0.45&    251&    0.962&           &      &	      &         &  4.782e+02&  3.11&    252&   0.941 \\
   HD37962 & 8.536e+02&  0.94&    252&  0.962&  5.469e+02&  0.60&    252&    0.963&           &      &	      &         &  1.963e+02&  3.20&    252&   0.955 \\
   HD38949*& 7.486e+02&  1.68&    250&  0.963&  4.843e+02&  1.58&    252&    0.969&           &      &	      &         &  1.722e+02&  4.73&    250&   0.961 \\
     P330E & 7.609e+00&  0.77&     62&  1.002&  4.994e+00&  0.31&      8&    1.025&  3.179e+00&  0.54&       5&    1.014&  1.747e+00&  4.33&      4&   0.997 \\
\enddata
\tablecomments{* Indicatates variability, see Section 5. Column 1 is the CALSPEC
name, Unc is the 1$\sigma$
uncertainty, N is the number of observations used, and Ratio is the ratio of
the IRAC Flux (mJy) to the CALSPEC synthetic photometry.}
\end{deluxetable}

\newpage
\section{New NLTE Model SEDs for OB Stars}		

For this paper, Hubeny has updated the OB NLTE model SED grids of
\citet{lanz03,lanz07} using a more complete hydrogen atom to produce more
realistic IR flux predictions. In particular, the new NLTE spectra have a much
better description of line confluences and of the contribution to the opacity in
very high series members. In addition, there are improvements in the treatment
of the hydrogen and He II continua from high levels, inclusion of more
scattering opacity sources, and updates to the physics of level dissolution.
Each model has 29957 points  at constant spectral resolution of R=5000 and a
micro-turbulent velocity of 2 km/s. The \citet{asplund09} element abundances are
used for computing the spectra.

The improvements in the physical description of hydrogen line profiles, level
dissolution, and the treatment of pseudo-continua influence the overall
atmospheric structure and the emergent spectrum. However, because the most
significant improvements occur for hydrogen lines in the infrared region, where
the flux for O and B stars is low, these changes do not cause significant
changes in the global model structure. Typically, the relative difference in the
continuum level amounts at most to a few times 0.1\%.

This finding permits a simplified computational strategy, using the model
atmosphere structure of temperature, electron density, and atomic level
populations from published
OSTAR2002{\footnote{http://tlusty.oca.eu/Tlusty2002/tlusty-frames-OS02.html}
and BSTAR2006{\footnote{http://tlusty.oca.eu/Tlusty2002/tlusty-frames-BS06.html}
grids to merely recompute emergent spectra with SYNSPEC for all available
models, which is several orders of magnitude less time-consuming than to
recompute full NLTE metal line-blanketed model atmospheres as was done
in \citet{lanz03,lanz07}.

See \citet{hubeny2021} for a description of the most recent versions of
\textsc{tlusty} and SYNSPEC. The new OB grid is available in the Mikulski
Archive for Space Telescopes (MAST) as a High Level Science Product via
\dataset[10.17909/bsbk-pj11]{\doi{10.17909/bsbk-pj11}}. The grid contain 1082
model pairs of spectra and continua and covers  effective temperature
($T_\mathrm{eff}$) in the range 15,000--55,000~K and surface gravity ($\log g$)
between 1.75 and 4.75, with steps of 0.25, where g has units of $cm~s^{-2}$. The
steps in $T_\mathrm{eff}$ are 1,000~K between 15,000 and 29,000~K and 2,500~K
between 30,000 and 55,000~K. As $T_\mathrm{eff}$ rises, more of the unstable
lower surface gravities are omitted. There are five metallicity values ($\log
Z$)=[M/H] of  0.301,  0.000, -0.301, -0.70, and -1.00, which correspond to Z=2,
1, 0.5, 0.2, and 0.1, where Z=1 is the solar value.

\newpage
\section{Variability and Rejected CALSPEC Stars}			

TESS data indicate variability in four CALSPEC standards \citep{mullally2022},
which vary by more than a peak-to-peak limit of 0.5\% and are flagged with an
asterisk in Table~\ref{table:irac_fluxes}. These stars and the amount of
variability (\%) in parentheses are 1732526 (1.40), 1808347 (1.65), 1812095
(1.57), and HD38949 (1.17). Such low levels of variability do not seriously 
affect our results, partly because of averaging over time and partly because
none of the four stars display anomalous flux levels. However, an extra  1.5\%
is included in their uncertainty estimates.

C26202 (a.k.a. 2MJ03323287) is faint and in a
crowded field, which makes its photometry too unreliable to include. WD0308-565
was considered for addition to the work of \citet{krick2021} but is rejected as
too faint to provide reliable IRAC photometry. LDS749B and WD1057+719 are even
fainter than WD0308-565 and are not reported here, even though LDS749B IRAC1
with a 4.6\% $1\sigma$ uncertainty happens to fall within 1\% of its model.
Sirius is an important CALSPEC star \citep{price04,bohlin2020}, but the Spitzer
Archive contains no observations which produce straightforward flux values.

\section{Results}			

\subsection{Comparisons among the Three Spectral Categories}

Figure~\ref{rcbcf} shows the ratio of our Spitzer IRAC photometry from
Equation~(\ref{eq5}) to the synthetic CALSPEC photometry from
Equation~(\ref{eql}) for our three categories of stars \citep{gordon2022}.
Purple symbols identify the IRAC photometry that has an uncertainty of $>$3\%.
In the top panel for our hot-star category, the three primary CALSPEC standards
(red) agree with IRAC within $2\sigma$, except for GD71 at 4.5 \micron, where
that IRAC2 is 2.6\%  ($3.9\sigma$) high. The IRAC3-4 data for the bright 10Lac
and $\lambda$ Lep with green connecting lines fall significantly below unity and
are purple because of the extra 3\% uncertainty assigned to these short 0.01 s
exposures, which have no available correction in Table~\ref{table:expcor} but
are likely to be systematically low, as suggested by the trends for IRAC3--4 in
Table~\ref{table:expcor}. The IRAC3 for $\eta$ UMa is 3.6\% high but that is
only $1.3\sigma$.

In the middle panel for the A-type stars, the ratios for all IRAC filters clump
tightly and suggest a rising trend from short to long wavelength. One
explanation for this trend could be that the CALSPEC models for A-stars are
systematically too dim as the wavelength increases from 3.6 to 8 \micron.
Perhaps, the LTE A-star models used for the CALSPEC SEDs should have more of a
rising slope from 3.6 to 8 \micron, as suggested by the NLTE discussion below
for the OB-stars. Alternatively, contamination by a range of contributions by
hot dust to the A-star IR fluxes seems unlikely, because the rms scatter among
the IRAC/CALSPEC A-star ratios from IRAC1 to IRAC4 does not increase above
1.3\% in Table~\ref{table:avg}.

\begin{figure}			
\includegraphics*[width=.95\textwidth,trim=0 35 1 35]{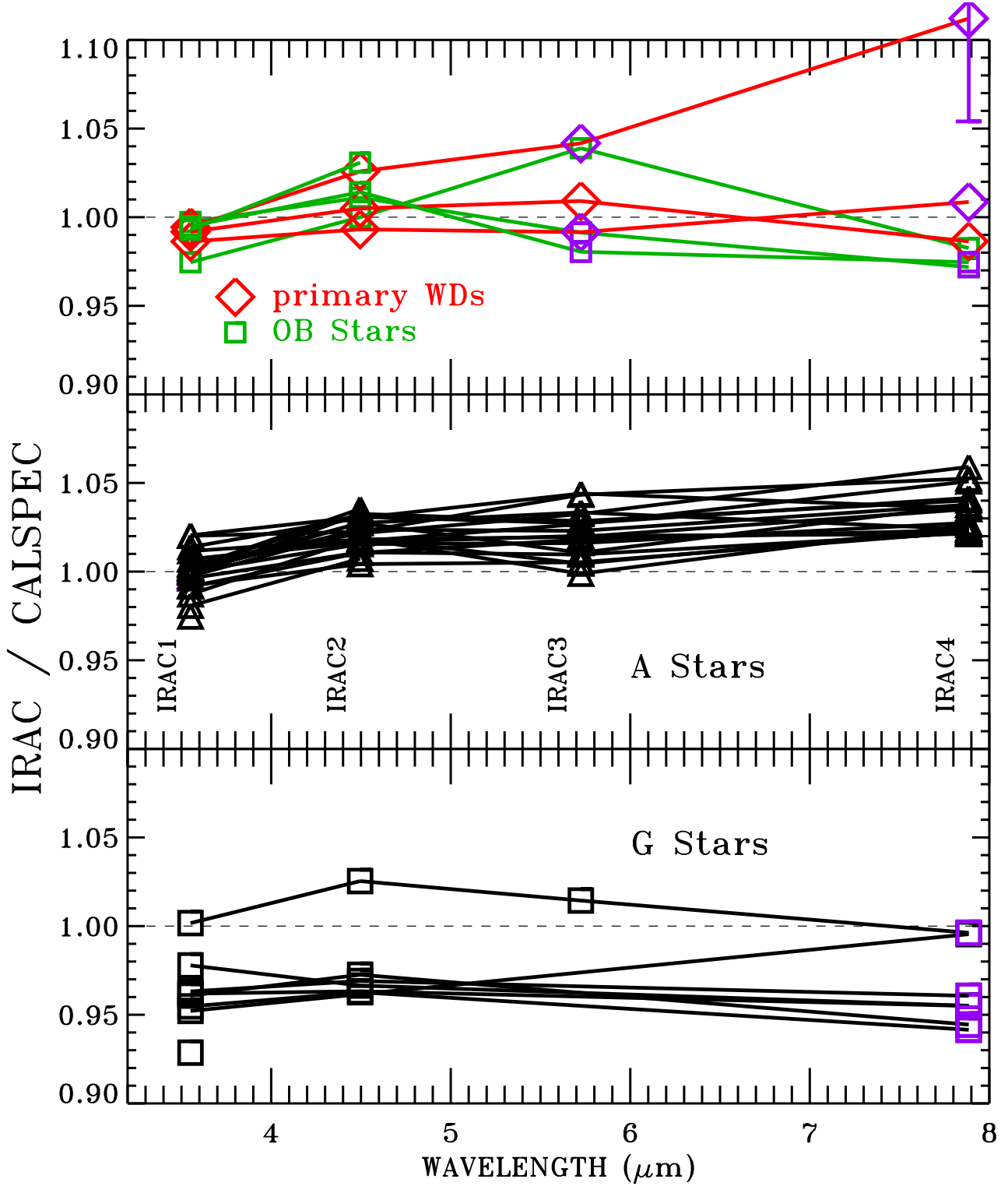}
\caption{\baselineskip=12pt
Ratio of photometry in the four Spitzer IRAC bands to synthetic photometry of
CALSPEC SEDs, where the synthetic photometry is from models normalized to the
STIS fluxes at 6800--7700~\AA. There are three panels for the three different
categories of JWST flux standards, i.e. the G-type stars, the A-stars, and the
hot WDs plus OB-stars. The model IR extentions of the OB CALSPEC standards are
updated with the new NLTE model grid. Purple symbols flag points with $>$3\%
IRAC uncertainties. A $1\sigma$ error bar appears for the most extreme point
that lies more than 10\% from unity to demonstrate that the ratio is still
within $2\sigma$ of unity.
\label{rcbcf}} \end{figure}

\begin{deluxetable}{lcccr}		
\tabletypesize{\normalsize}
\tablewidth{0pt}
\tablecolumns{5}
\tablecaption{\label{table:avg} IRAC / CALSPEC}
\tablehead{
\colhead{Channel} &\colhead{Star Set} &\colhead{IRAC/CALSPEC} 
	&\colhead{rms} &\colhead{N}}
\startdata
IRAC1  & Prime WDs  &0.991 &0.004  & 3 \\
IRAC2  & Prime WDs  &1.005 &0.016  & 3 \\
IRAC3  & Prime WDs  &1.012 &0.025  & 3 \\
IRAC4  & Prime WDs  &1.010 &0.066  & 3 \\
& \\
IRAC1  & A          &1.000 &0.011  &22 \\
IRAC2  & A          &1.022 &0.009  &21 \\
IRAC3  & A          &1.018 &0.013  &17 \\
IRAC4  & A          &1.034 &0.011  &17 \\
& \\
IRAC1  & G	  &0.963 &0.022  & 8 \\
IRAC2  & G	  &0.978 &0.025  & 6 \\
IRAC3  & G	  &1.014 &	& 1 \\
IRAC4  & G	  &0.991 &0.023  & 7 \\
\enddata
\end{deluxetable}

In the bottom panel for the G stars, there is only one IRAC3 measure and most of
the IRAC4 points have large uncertainty, so not much can be concluded regarding
the averages for IRAC3-4. The five low values of the seven IRAC4 points have
have only 0.01 s exposures, so are probably systematically low like the O-stars
10Lac and  $\lambda$ Lep, while 16CygB and P3303 are near unity with their
longer exposure times of 1 and 26.8 s, respectively. While the main clump of
IRAC1-2 G-star points are robust and low by 2--7\%, P330E lies high by 3--4\%
with uncertainties as low as $1\sigma$=0.31\% for IRAC2. The P330E ratios with
exposure times of 10.4s are consistent with the A-star locus, while all 12 of
the low IRAC1--2 G-star points are entirely from 0.01 s exposures. However, for
stars hotter than G, there are 10 IRAC1--2 photometry values dominated by 0.01 s
exposures; and none of these 10 have IRAC fluxes as low as the G-star
IRAC/CALSPEC weighted averages of 0.963 and 0.978 for IRAC1 and IRAC2,
respectively, from Table~\ref{table:avg}. In particular, the hotter star 0.01 s
ratios range from 0.981--1.001 for IRAC1 and 1.007--1.033 for IRAC3.
Furthermore, non-linearities cannot cause the G-stars to have low fluxes,
because the detector full-wells are filled to similar depths of a few percent
for all 22 of the 0.01 s exposure times from all three categories of stellar
type.

Thus, the possibility seems unlikely that all of the G-star IRAC1--2 0.01
exposures are erroneous. A likely explanation for the low IRAC/CALSPEC ratios
for many G-stars is an inadequate STIS spectral range of only 0.3--1 \micron\
for fitting models, while P330E has WFC3 and NICMOS data extending to 2.5
\micron\ to better constrain its fit. In figure 4 of \citet{bohlin2017}, the
fits in the IRAC wavelength range to STIS alone for G-stars tend to lie above
the fits to STIS+NICMOS by up to $\approx$3\% for P330E. A WFC3 IR grism
campaign to measure the 1--1.7 \micron\ SEDs of the seven deficient G-stars
should be done to
provide more reliable models for predicting fluxes longward of 2 \micron.


Table~\ref{table:avg} summarizes the results, where the hot-star category
includes only the three primary WDs. Small offset in IRAC/CALSPEC fluxes for any
group of stars is not surprising, as the uncertainties of both the CALSPEC and
the IRAC flux calibrations are of order 2\%. However, the systematically rising
offsets from zero at 3.6 \micron\ to 3\% at 8 \micron\  for the A-star group
differ from the hot-star group. That systematic difference cannot be caused by
an IRAC flux calibration error, as the same IRAC calibration is applied in the
same way for all three groups. To make the A-star models agree with the IRAC
data, the required  amount of brightening at 8 \micron\ relative to 3.6 \micron\
is $\approx$3\% for the LTE BOSZ models \citep{bohlin2017} used to extend the
HST/CALSPEC fluxes to IRAC wavelengths. Alternatively, the IRAC data for the
three faint primary WDs relative to the generally brighter ensemble of A-stars
is in error; or the well-vetted models for the primary WDs \citep{bohlin2020}
are erroneous. However, there is a small possibility that the WD/A-star offsets
are mostly  statistical, even though all four WD ratios lie below the
corresponding weighted average A-star ratio. Non-linearity seems an unlikely
cause of the WD vs. A-star problem, because the brightest WD G191B2B has nearly
the same flux level as the faintest A-star 1743045. None of the four G191B2B
IRAC photometry results exceeds its model by more than 0.6\% but 1743045 is
among the highest A-stars at 3--6\% above its LTE BOSZ model.

\subsection{Comparisons with the SEDs of the Hot Stars}		

Figure~\ref{sed} compares the IRAC photometry to CALSPEC SEDs for the hot stars
in the top panel of Figure~\ref{rcbcf}, where the IRAC data (triangles) are
plotted at their pivot-wavelengths and their deviations from the red and green
SEDs are the same as the deviation from unity in Figure~\ref{rcbcf}.

While the blue curves are the LTE BOSZ models fit to the STIS data at 6800--7700
\AA\ for the three OB-stars and are the current versions in CALSPEC, the green
curves are the same fits but using our new NLTE OB-star grid and have similar
reduced $\chi^2$. The biggest difference in reduced $\chi^2$ is for 10 Lac with
4.0 and 2.8 for the NLTE and BOSZ residuals, respectively. Because the three
IRAC1 points are closer to the green NLTE models and all three IRAC2 points are
much closer to the NLTE models, the green NLTE models are a much better match to
the reliable IRAC1--2 fluxes; and NLTE versions stis\_006 and will replace the
stis\_005 versions of LTE models in CALSPEC. Hot WD models show a similar effect
for the IR flux of the G191B2B NLTE model, which rises from agreement with an
LTE model at 1 \micron\ to $\approx$10\% higher at 8 \micron\ in figure 3 of
\citet{bohlin11}.

\begin{figure}			
\centering 
\includegraphics*[width=.96\textwidth,trim=60 50 0 75]{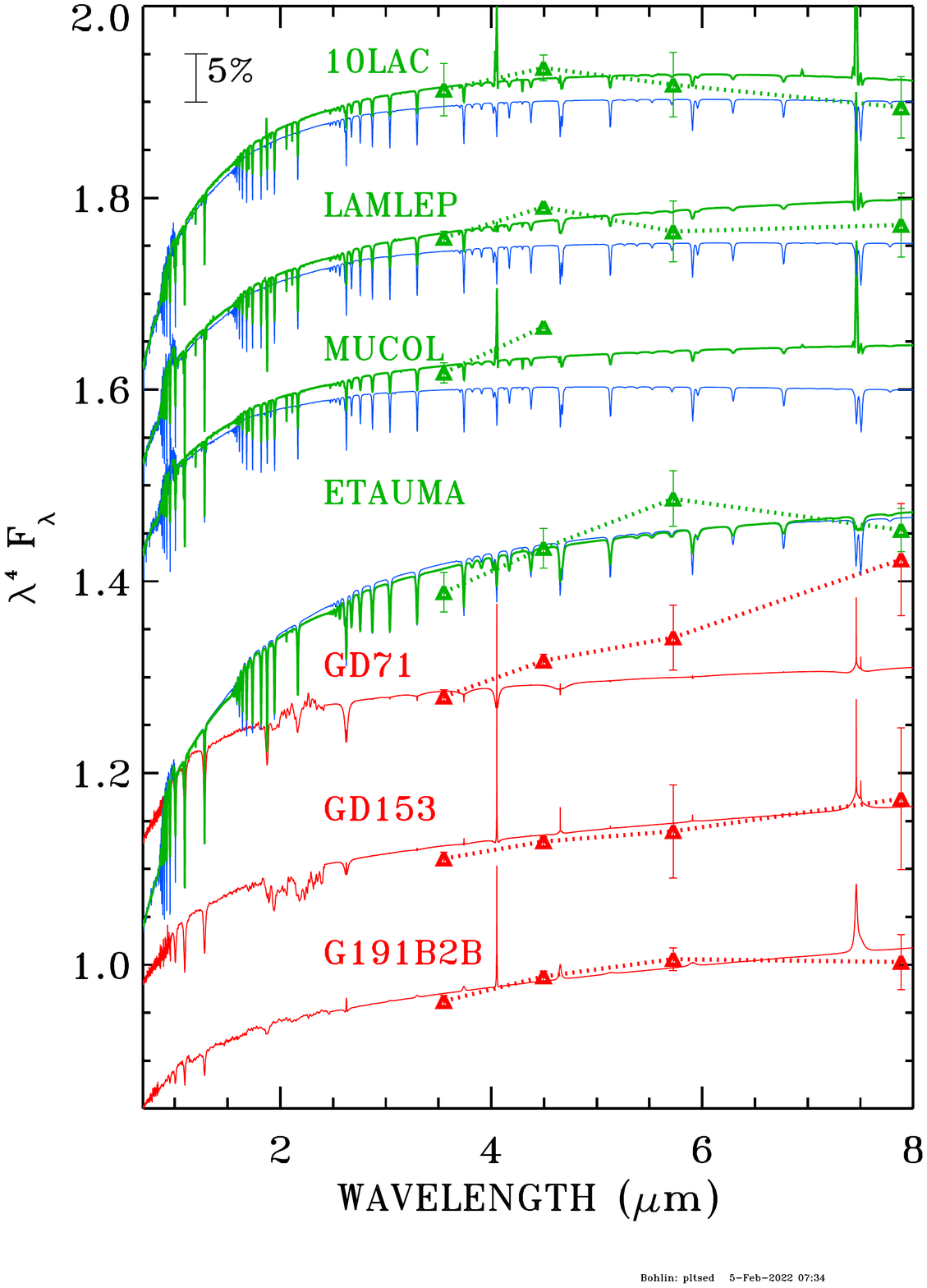}
\caption{\baselineskip=12pt
The SEDs scaled by $\lambda^{4}$ and then scaled to unity near 6 \micron\ for
the seven stars in the top panel of Figure~\ref{rcbcf}, using the LTE models for
OB-stars (blue). Each star above G191B2B is offset by 0.15 in Y. The green SEDs
are the new NLTE models matched to the blue LTE models at 6800--7700~\AA. The
red SEDs are the three primary WDs from CALSPEC, as specified in
\citet{bohlin2020}. Triangles illustrate the difference between our IRAC
photometry and CALSPEC at the four IRAC pivot wavelengths. The most reliable
IRAC1-2 photometry agrees much better with the NLTE models (green) than with the
LTE models (blue) for the three stars at the top.
\label{sed}} \end{figure}

\subsection{Absolute Flux Calibration on the CALSPEC Scale}		

Table~\ref{table:avg} summarizes the weighted average results for the three
categories of stars, where only the three fundamental primary WDs are included
for the hot star  category. These three primary WDs form the basis of the
HST/CALSPEC absolute flux system; and the agreement with the existing IRAC flux
calibration is, perhaps, fortuitous because of our adjustments to the subarray
aperture corrections, our adjustments to the IRAC sensitivity changes with time
in Table~\ref{table:tcor}, and our revision of the exposure times in
Table~\ref{table:expcor}. However, these differences are minor in many cases,
e.g. our flux values in Table~\ref{table:irac_fluxes} agree to a maximum
difference of 3\% for G191B2B IRAC4 among the eight measures of the three prime
WDs reported by \citet{krick2021}. Furthermore, the good agreement of the four
OB-stars in Figure~\ref{sed} with their green model curves reinforces the WD
results, especially for IRAC1--2.

The Table~\ref{table:avg} results for the three prime WDs demonstate agreement
between the IRAC and CALSPEC fluxes within 1\% for all four IRAC channels when
using the existing flux calibration of \citet{carey2012}. The most significant
difference is 0.991, i.e. 0.9\%, with an rms uncertainty of 0.004, i.e. 0.4\%.
This 0.9\% offset is only 2.3$\sigma$ and does not justify a revision to the
existing IRAC flux calibration. 

\subsection{Comparison with the \citet{krick2021} Fluxes}	

There is good agreement of our Table~\ref{table:irac_fluxes} with the eight
\citet{krick2021} values for the three primary WDs, but Krick does not report 
IRAC3--4 values for GD153 or GD71. Again in the hot star category, the results
are similar for the three OB stars 10 Lac, $\lambda$ Lep, and $\mu$ Col, while
Krick does not include the saturated $\eta$ UMa.  For the A-stars, the biggest
difference with our Table~\ref{table:avg} is 2\% for  IRAC1 for 18 stars, while
our total star count is 22. For the G-stars, Krick gets better agreement with
the models for IRAC1--2, but  worse agreement and larger scatter for IRAC4. For
P330E, both sets of average fluxes agree to $<$2\%, except for IRAC4, where the
Krick value is 9\% larger than our Table~\ref{table:irac_fluxes} value that
agrees with the P330E model. The set of eight G-stars is the same for both
analyses.

\subsection{Fitting Models with IRAC Fluxes}	

Could the overall model fits be improved by adding the IRAC constraints to the
HST fluxes and then refitting to get new models? For the A-stars, the new fits
change by no more than $\approx$1\%, which validates the  fitting to the HST
SEDs only. However, for the G-stars, major changes result with the discovery
that there are models that fit both the HST and IRAC fluxes nearly perfectly,
i.e. typically within 1\% on average. These fits seem wonderful at first glance,
but upon close examination, the resulting reddening, E(B-V) extinctions, are
unreasonably large. For example, the G2V star 18 Sco has a (B-V)=0.65 for an
expected E(B-V)=0.02 mag. However, the model fit to the STIS+IRAC fluxes
requires E(B-V)=0.10, i.e. a 0.08 mag discrepancy compared to the usual expected
photometry errors of 0.01--0.02 mag. A resolution of this problem will come 
from JWST measures of A-star/G-star brightness ratios. If these JWST ratios
agree with both of the same ratios for STIS in the 0.8--1 \micron\ range and for
IRAC, then the insufficiently constrained G-star models must have the
wrong shape over this wavelength range. Because of this G-star quandary,
achieving the small gains  from using the IRAC data to constrain just the hotter
star models seems to be an unworthy exercise.

\section{Conclusions}		

Table~\ref{table:avg} demonstrates agreement to $\approx$1\% between IRAC and
CALSPEC for the three primary WD standard stars G191B2B, GD153, and GD71.
Similarly, the average differences for four hot OB stars confirm the WD results
to 2\% for IRAC1--3. The average results for the robust set of 17--22 A-stars 
show agreement with the hotter star results for IRAC1 but are discrepant by as
much as 3.4\% at IRAC4. In the 4.5--8 \micron\ wavelength region, the 2--3\%
A-star offsets from the WD ratios of IRAC/CALSPEC define the uncertainties of
the current CALSPEC hot star SEDs with respect to the A-star category. For the
G-star category, the set of data for IRAC3--4 lacks sufficient precision, while
most of the IRAC1--2 G-star ratios to CALSPEC are discrepant with the hotter
star ratios. However, one star P330E with WFC3 spectrophotometry to 1.7 \micron\
and  NICMOS data to 2.5 \micron\ agrees with the A-star results. The other seven
G-stars have only STIS data that exend to just 1 \micron. When fitting P330E
using just the STIS data, the fitted model in the IRAC wavelength range is
$\approx$3\% above the model fitted to the full range to 2.5 \micron. Thus, a
major contributor to the IRAC/CALSPEC G-star discrepancy is inadequate spectral
coverage for fitting models, which makes the models systematically too bright
by up to 4\% in  Table~\ref{table:avg}.
 The G-star STIS spectra cover only 0.3--1 \micron\ in
comparison to a minimum range of 0.17--1 \micron\ for hotter stars. Therefore,
the current CALSPEC model extensions for the seven G-stars with little flux
below 0.3 \micron\ and no constraining HST data longward of 1 \micron\ are
likely to be a few percent too high in the IRAC wavelength range.

In summary, there is no strong evidence from the IRAC data for problems
with the spectral grids used for predicting the IR SEDs. Inadequacies of the
observational data both from HST and Spitzer contribute to deviation of the models
from the IRAC photometry. To be conservative, uncertainties of 2, 3, 4\%
longward of 3 \micron\ should be adopted for our hot-star, A star, and G star
categories, respectively. JWST measures of hot-stars with respect to the A and G
stars can be used to correct the cooler star models and reduce their uncertainty
relative to the more accurate hot-star category.



\begin{acknowledgments}

Support for this work was provided by NASA through the Space Telescope Science
Institute, which is operated by AURA, Inc., under NASA contract NAS5-26555. This
work is based on observations made with the Spitzer Space Telescope, which was
operated by the Jet Propulsion Laboratory, California Institute of Technology
under a contract with NASA. \end{acknowledgments}

~\\
\textbf{ORCID IDs}

Ralph C. Bohlin https://orcid.org/0000-0001-9806-0551

Jessica E. Krick https://orcid.org/0000-0002-2413-5976

Karl D. Gordon https://orcid.org/0000-0001-5340-6774

Ivan Hubeny https://orcid.org/0000-0001-8816-236X

\bibliographystyle{/Users/bohlin/pub/apj}

\bibliography{/Users/bohlin/pub/paper-bibliog.bib}

\end{document}